\documentclass[a4,twocolumn,12pt]{article}
\usepackage{amssymb}
\usepackage[dvips]{epsfig}
\usepackage[dvips]{epsfig}

\begin{document}

\title{Another Co*cryption Method}

\author{Bruno Martin\footnote{\texttt{Bruno.Martin@unice.fr}}\\
{\small Universit\'e de Nice Sophia-Antipolis,
Laboratoire \textsf{I3S}, Projet Recif, UMR CNRS 6070,}\\ 
{\small 930 Route des Colles, BP 145,
F--06903 Sophia--Antipolis Cedex, France.}
}
\date{}

\maketitle

\begin{abstract}
  We consider the enciphering of a data stream while being compressed
  by a LZ algorithm. This has to be compared to the classical
  encryption after compression methods used in security
  protocols. Actually, most cryptanalysis techniques exploit patterns
  found in the plaintext to crack the cipher; compression techniques
  reduce these attacks. Our scheme is based on a LZ compression in
  which a Vernam cipher has been added. We make some security remarks
  by trying to measure its randomness with statistical tests.  Such a
  scheme could be employed to increase the speed of security protocols
  and to decrease the computing power for mobile devices.

  \emph{\small Cryptography, compression, pseudo-random sequences, security.}
\end{abstract}
\section*{Introduction}

Information security is currently one of the main challenges in
computer networks. In the emergent communication paradigm where
wireless and wired networks are interoperating, security issues become
crucial. Traditional technologies are every day more inadequate and
existing standards should be improved for use in resource restricted
environments. We aim to develop a secure algorithm for
confidentiality, but cheaper in terms of size and computing power.
 
In many security protocols, a compression algorithm is run
prior encrypting the data to increase the security and
the bandwidth. These algorithms are run on the original stream. They
all stem from research by J.~Ziv and A.~Lempel who have designed two
compression algorithms: LZ77 and LZ78 \cite{salomon}. After
compression, if the speed of computation is taken into account, the
compressed data is enciphered with the use of a stream cipher like
RC4 (let us recall that RC4 is 15 times quicker than a 3DES and is
used in protocols like WEP and SSL~\cite{rescorla}).

In the present paper, we propose to scramble (encipher) a data stream
while it is being compressed. We assume the reader familiar with
classical compression algorithms and with secret key cryptography for
which a good introduction is~\cite{stallings}.  The paper is organized
as follows: section~\ref{sec:idea} presents the basis of our idea,
while section~\ref{sec:related} recalls the related results. In
section~\ref{sec:toy-scheme} we illustrate our idea with a ``toy''
implementation which uses a compression algorithm from the Lempel-Ziv
family. Some statistical tests have been made and are presented in
section~\ref{sec:analysis}.

\section{The idea}
\label{sec:idea}

Our idea is to encipher the data stream while it is being compressed
by a lossless dictionary algorithm. The basic idea which motivates
this proposition is that a compressed stream is already almost random
and a good candidate to be scrambled by a simple Vernam cipher. This
comes from the notion of incompressibility introduced with Kolmogorov
complexity. A.N.~Kolmogorov~\cite{kolmogorov} has proposed a
complexity which speaks about objects rather than the usual classes of
languages addressed by classical complexity. Informally, Kolmogorov
complexity corresponds to the size of the smallest program $p$ which
can print out on its standard output the object $x$. If
$\sharp p<\sharp x$, we say that $x$ is \emph{compressible}, otherwise
\emph{incompressible}. It provides a modern notion of randomness
dealing with the quantity of information in individual objects which
says that an object $x$ is random if it cannot be represented by a
shorter program $p$ whose output is $x$ or, in other words, if $x$ is
incompressible~\cite{salomon}. From this point of view, the output of
any compression algorithm is an approximation of a random sequence,
although highly reversible. Our idea is to scramble the output of a
compression algorithm by a Vernam cipher and to do this while
the data stream is being compressed in order to avoid to pass the
compressed data stream to another encryption process. From the above
discussion, the output of our scheme should be almost random.

\section{Related work}
\label{sec:related}

Actually, there are two methods sharing the same idea but in a slightly
different way. The first one is called \emph{concryption} and has been
patented by Security Dynamics (US Patent \#5479512). It is a
method for the integrated compression and encryption (concryption) of
clear data. For concryption, the clear data and an encryption key are
obtained, at least one compression step is performed and at least one
encryption step is performed utilizing the encryption key. The
encryption step is preferably performed on the final or intermediate
results of a compression step, with compression being a multistep
operation.  The second method is called
\emph{compryption}~\cite{comcryption-rsa} and is due to R.~E.~Crandall
when he was Apple's Chief Cryptographer. Roughly, his idea is to index
a great number of entropy compression algorithms by a secret key. He
then gets a holistic (one-pass) compress/encrypt algorithm.  This
method is currently used for enciphering the passwords in the keychain
application starting with Mac OS 9 and still used in Mac OS X from
Apple. It is recorded under US Patent \#6154542, ``Method and apparatus for
simultaneously encrypting and compressing data''.

\section{The proposed scheme}
\label{sec:toy-scheme}
We use the mode of operation of LZ 78 which uses a growing
dictionary~\cite{salomon}.  It starts with $2^9=512$ entries (with the
first 256 entries already filled up, eventually after an initial
permutation). While this dictionary is in use, 9 bit pointers are
written onto the output stream after encryption by a Vernam cipher.
When the original dictionary is filled up, its size is doubled to 1024
entries and 10 bit pointers are then used (and encrypted as well)
until the pointer size reaches a maximum value set by the user. When
the large dictionary is filled up, the program continues without
changes to the dictionary but with monitoring the compression ratio.
If this ratio falls down a predefined threshold, the dictionary is
deleted and a new 512 entries dictionary is started. The algorithm
below presents the scheme. In the sequel, we denote by PRBS a
\emph{pseudo-random Boolean sequence}.

{\small 
\begin{verbatim}
Index = 256; Length = 9; Word = null; 
Limit = 12;
Initialise 256 inputs in Dictionary 
//(eventually after a permutation)
//(a+b stands for concatenation)
REPEAT 
  read S
  //(Read a symbol from the stream)
  IF Word+S is in Dictionary 
     THEN 
       Word = Word+S; 
       Emit = false
     ELSE
       Output(index of Word) XOR (PRBS)  
       // Vernam cipher
       Index of (Word+S) = Index; 
       Index++
       IF Length =  Limit 
         THEN
             Re-initialise Dictionary 
       ENDIF
       IF Index = 2Length 
         THEN Length++ 
       ENDIF
       Word = S; Emit = true 
  ENDIF
UNTIL no data found 
IF Emit =  false THEN
  Output the (index of Word) XOR (PRBS) 
  // Vernam cipher
ENDIF 
\end{verbatim}
}

The implementation was just made as a proof-of-concept in \texttt{C++}
and using the LEDA library~\footnote{Avaible from \texttt{http://www.mpi-sb.mpg.de/LEDA}.} which provides a sizable
collection of data types and algorithms in a form which allows them to
be used by non-experts. 

The difference with concryption is that we use a single pass
compression algorithm while they require the compression to be a
multistep operation, and it is not based on an entropy compression
algorithm used in the comcryption method, although an entropy
compression algorithm could be added to shorten the mostly used
pointers which are returned by the algorithm.
\section{Analysis}
\label{sec:analysis}

\begin{figure}[htbp]
  \centerline{\epsfxsize=8cm\epsfbox{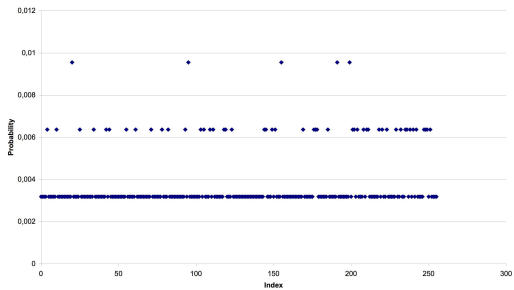}}
  \vspace*{-1cm}
  \caption{Probabilities of the output.}
  \label{fig:sortie}
\end{figure}

Though a plot of the output (see figure~\ref{fig:sortie}) is rather
encouraging, we were deceived while testing outputs with a $\chi^2$
test for which the results were a little bit too weak.  This may be
explained by the rather bad choice of a ``toy'' linear feedback shift
register for generating the PRBS. We expect the result to be improved
with the use of good pseudo-random generators like RC4, or even with
so-called ``perfect'' PRBS generators.

Further testing should be made according to~\cite{FIPS} which requires
a PRG to pass a number of statistical tests or the
Marsaglia tests, a set of 23 very strong tests of randomness
implemented in the Diehard program\footnote{Avaible from
  \texttt{http://diehard.darwinports.com}.}.

\section{Discussion}

Although not truly pseudo-random (but this is also not a pseudo-random
generator), the output of our compression and encryption scheme is
encouraging if we look at the typical output depicted by
figure~\ref{fig:sortie}. Further study should be made with the help of
a good pseudo-random generator with classical tests and a fine tuning
of all the parameters.

The use of compression and encryption mixed together should increase
the bandwidth, decrease the latency as well as it also might decrease
the energy consumption required for the same purpose when using
encryption after compression for mobile devices or RFID. 


\bibliographystyle{plain}
\bibliography{enseigne}

\begin{thebibliography}{1}

\bibitem{comcryption-rsa}
R.E. Crandall.
\newblock Comcryption.
\newblock In {\em RSA Data Security Conference}, 1998.

\bibitem{kolmogorov}
A.N. Kolmogorov.
\newblock Three approaches to the quantitative definition of information.
\newblock {\em Problemy Pederachi Informatsii}, 1:3--11, 1965.

\bibitem{FIPS}
National~Institute of~Standards~Technology.
\newblock {FIPS} publication 140-2, security requirements for cryptographic
  modules.
\newblock US Gov. Printing Office, Washington, 1997.

\bibitem{rescorla}
E.~Rescorla.
\newblock {\em {SSL} and {TLS}: Designing and Building Secure Systems}.
\newblock Addison-Wesley, Reading MA, 2001.

\bibitem{salomon}
D.~Salomon.
\newblock {\em Data compression, the complete reference}.
\newblock Springer Verlag, 1998.

\bibitem{stallings}
W.~Stallings.
\newblock {\em Cryptography and Network Security}.
\newblock Prentice-Hall, 4th. edition, 2006.

\end{thebibliography}
\end{document}